\documentclass[prd,groupedaddress,nofootinbib,showpacs,preprintnumbers,floatfix]{revtex4}

\usepackage{mathrsfs}
\usepackage{amsfonts,amssymb,amsmath}
\usepackage{feynmp}
\usepackage{graphicx}
\usepackage{multirow}
\usepackage{slashed}
\usepackage{fancybox}
\usepackage{subdepth}
\usepackage{float}
\usepackage{fancyvrb}
\usepackage{accents}
\usepackage{stmaryrd}
\usepackage{wasysym}
\usepackage{comment}
\usepackage[T1]{fontenc}
\usepackage{etoolbox}
\usepackage{bm}
\usepackage{url}
\usepackage{hyperref}
\usepackage{breakurl}
\usepackage{ragged2e}
\usepackage{lipsum}
\usepackage[nodisplayskipstretch]{setspace}
\usepackage{color}

\DeclareMathSymbol{\widetildesym}{\mathord}{largesymbols}{"65}

\begin{document}

\title{Entanglement in the ground state of clusters joined by a single bond}
\author{Barry Friedman and Alyssa Horne}
\affiliation{
	\mbox{Department of Physics, Sam Houston State University, Huntsville, TX 77341-2267, USA}\\}


\begin{abstract}
The ground state of an antiferromagnetic Heisenberg model on $L \times L$ clusters joined by a single bond and balanced Bethe clusters are  investigated with quantum Monte Carlo and modified spin-wave theory.  The improved Monte Carlo method of Sandvik and Evertz is used and the observables include valence bond and loop valence bond observables introduced by Lin and Sandvik as well as the valence bond entropy and the second Renyi entropy.  For the bisecting of the Bethe cluster, in disagreement with our previous results and in agreement with modified spin-wave theory, the valence loop entropy and the second Renyi entropy scale as the $\log$ of the number of sites in the cluster.  For bisecting the $L \times L - L \times L$ clusters, the valence bond entropy scales as $L$, however, the loop entropy and the entanglement entropy scale as $\ln(L)$.  For the entanglement entropy,   the coefficient of the logarithm, the number of Goldstone modes/2,  is universal and is the analogue of the sub leading term of a $L \times L$ cluster.   This result was substantiated by a modified spin-wave theory calculation for the ferromagnetic $X-Y$ model.  Taken together,  the calculations suggest that linking high entanglement objects will not generate much more entanglement. As a consequence, simulating the ground state of clusters joined together by a few bonds should not be much more difficult than simulating the ground state of a single cluster.  \end{abstract}


\pacs{03.67.Mn, 75.10.Jm, 75.40.Mg}


\maketitle


\section{Introduction}

\begin{figure}
 \includegraphics[width=0.60\textwidth]{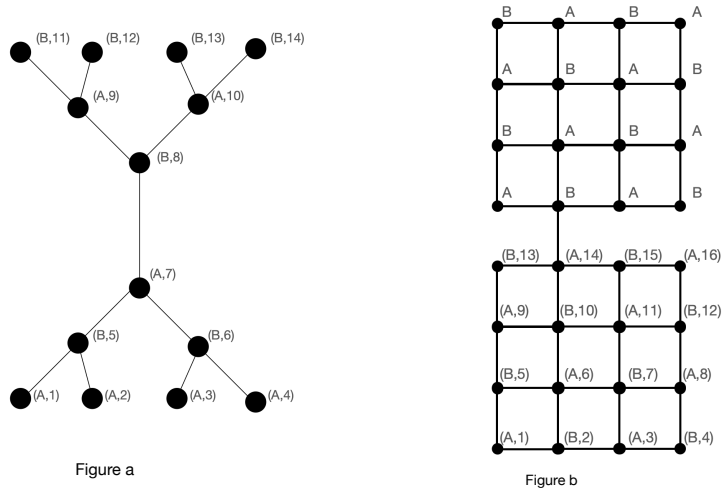}
\caption{   \label{FIG:CLUSTER}Figure a :  14 site three branch Bethe cluster.   Figure b : $4 \times 4 - 4 \times 4$ cluster The pair (A(B), i) refers to the ith site and the A(B) sub lattice. }
\end{figure}

This paper is an investigation of the ground state quantum mechanical entanglement between two clusters joined by a single bond. (See figure \ref{FIG:CLUSTER}).  The original motivation was two-fold.  First of all, numerically, it has been observed that it is much harder to simulate Heisenberg models on Bethe clusters \cite{otsuka,friedman1,kumar,chan1} than on one dimensional chains.  If there is high entanglement entropy in bisecting Bethe clusters, this would explain this observation.  Secondly, it has been suggested \cite{levine} by studying clusters of non interacting fermions joined by a single link one could better understand the area law in two dimensions.  In particular, the $\ln L$ dependence of one dimension, goes over to $L \ln L$ if each link adds a term $\ln L $.  

Further calculations \cite{ caravan} motivated by the above considerations led to surprises.  Firstly, for an isotropic antiferromagnetic Heisenberg model on a balanced Bethe cluster, the valence bond entropy scales as the number of sites in the cluster (see appendix A0 for a discussion).  Secondly, for non interacting fermions, at half filling,  on a $L \times L - L \times L$  cluster  (2 $L \times L$ clusters joined by single hopping term) with $L$ even, the entanglement entropy, $S_1$, scales as $L$.  (See appendix A0 for a discussion for similar result for the Bethe cluster).   These results are surprising due to the area law \cite{eisert} for the entanglement entropy of the ground state.  

Recall that the area law states that the entanglement entropy in the ground state scales as the size of the boundary, not the volume of a subsystem.  The area law  is expected to hold when there is a gap between the ground state and the first excited state and is shown to be true in a number of physical examples.  In particular, there is a mathematical proof for one dimensional gapped systems \cite{hastings1,arad}    For gapless models, non interacting fermions and gapless systems in one dimension, examples show the possibilities of additional multiplicative log terms.  For a gapless model, with stronger size dependence  for the entanglement entropy, see \cite{movassagh}.   \textcolor{red} {There are also non translationally invariant examples with spatially dependent Heisenberg interactions having volume law scaling \cite{ramirez}.} 

Thus one might expect for bisecting the Bethe Cluster and $L \times L - L \times L$ clusters studied in \cite{caravan}  that at most one would get a $\ln N$ scaling with the number of sites $N$ since there is only one site joining the subsystem to the rest of the cluster.   To further investigate the apparent stronger size dependence,  a Monte Carlo method \cite{sandvik,hastings2,kallin2,kallin3} was used to study the antiferromagnetic Heisenberg model  on the  Bethe Cluster and $L \times L - L \times L$ clusters \cite{friedman}.  By using the Monte Carlo method, it was found that $S_2$, the second Renyi entropy for the Bethe cluster scaled as $N$ and the valence bond entropy for the $L \times  L - L \times L$ cluster scaled as $L$ consistent with the results of \cite{caravan}.

However,  the results in \cite{friedman} were limited in system sizes and number of projections used in the Monte Carlo process (see Appendix A ).   To critically examine the results of \cite{friedman}   two approaches have been undertaken, firstly the use of better observables and secondly by employing a better numerical method.  First of all, Sandvik and Lin \cite{lin} have  proposed two quantities  related to the valence bond entropy that are easier to calculate than the Renyi entropies but have closer correspondence to the Renyi entropies than the valence bond entropy.  One which will be referred to as $<n_S>$ counts the valence bonds joining a subsystem to its complement, the second referred to as $<l_S>$  counts the loops generated by  the double projector Monte Carlo process.  These quantities were introduced as an easier way to compute a quantity that has a greater correspondence to the second Renyi entropy and the entanglement entropy.   For a brief discussion of the loop entropy see appendix B.  Both these quantities can be calculated  using the improved Monte Carlo technique of Sandvik and Evertz \cite{evertz,lin}.   The algorithm uses a mixed valence bond, z basis (i.e. the ordinary computational basis) with loop updates.  This method can also be used to calculate the valence bond entropy and even $S_2$ \cite{kallin3}, \cite{kallin4}. \textcolor{red} {An explanation of the algorithm and its relation to observables is given in appendix A. }

The paper is organized as follows.  In the following section, entanglement in  the ground state of antiferromagnetic $L \times L - L \times L$ clusters is investigated by the improved Monte Carlo method and better observables.  In the second section, the Bethe clusters, are studied by the improved Monte Carlo methods and modified spin-wave theory.  The third section returns to the $L \times L - L \times L$ cluster using modified spin wave theory to treat larger systems.  This section also includes calculations for entanglement in the ground state of the ferromagnetic $X-Y$ model using modified spin wave theory.  The final section consists of a summary and conclusions.

\section {$L \times L$ clusters joined by a single bond}

The first quantity studied is the valence bond entropy for 2  $L \times L$ clusters joined by a single bond and the bond is chosen to be the closest bond to the middle of the side between the two clusters.  The Heisenberg Hamiltonian considered has 
$J >  0$ with nearest neighbor interactions only and the spin operators have spin $1/2$ 
\begin{equation}
H =  J \sum_{<i,j>} \bf{S_ i} \cdot \bf{S_j}
\end {equation} 
where  $<i,j>$ refer to nearest neighbors.

All calculations done in this paper are for the ground state with free boundary conditions. Free boundary conditions are more natural for density matrix renormalization group (dmrg) calculations \cite{white}.   \textcolor{red}{If periodic boundary conditions are used between the ends  of the two $L \times L $ clusters, bisecting the   system would give an entropy proportional to the $L$ sites connecting the two ends of the clusters.  This is not the case for periodic boundary conditions across each cluster, i.e. two tori joined by a single bond.}

Recall that even for a $L \times L$ cluster there is no exact solution and there is no proof  of long range antiferromagnetic order for the ground state of the spin $1/2$ model \cite{tasaki}.  However, since there is no sign problem, Monte Carlo is an effective numerical method, large systems can be accessed and there is good numerical evidence there is long range antiferromagnetic order \cite{liang}.  Physically, the model is relevant for the undoped state of the cuprate superconductors.  We also note, that  the $\bf{S_ i} \cdot \bf{S_j}$
operator for spin $1/2$ can be written as the exchange of the z component of spin \cite{feynman}, hence this operator has a natural interpretation in computer science as bit exchange.

\begin{figure}
\includegraphics[width=0.30\textwidth]{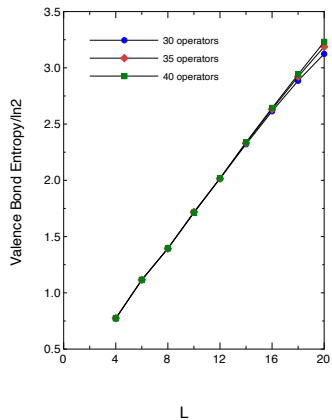}
\caption{ \label{FIG:VB1} Valence Bond entropy vs. $L$ for $L \times L-L \times L$ clusters. The blue circles are for projections with 30 operators, the red diamonds 35 operators and the green squares 40.  The lines are guides for the eye.  Statistical errors are smaller than the symbols in the figure.   }
\end{figure}

To calculate the valence bond entropy the loop algorithm is used (see \cite{evertz} figure 4 and appendix A). The results are shown in figure \ref{FIG:VB1}.  The calculations are done for $L$ even to avoid an even-odd effect, statistical errors are smaller than the symbols in the figure.  Up to $L=20$  there appears to be good linearity in the valence bond entropy with better linearity when more operators are included in the projection; projections with $2 \times 30$ $L ^2$, $2 \times 35$ $L^2$,  $2 \times 40$ $L^2$ operators were checked.  This is single projector Monte Carlo, $2 \times L^2$ is the system size \cite{evertz} \cite{kallin2}.  All the results were obtained with a starting valence bond state for which nearest neighbor horizontal sites are joined by a valence bond i.e. figure 2 (A,1, B,2) (A,3 B,4) (B,5 A,6) (B,7 A,8) etc..  For this starting configuration there are no valence bonds joining the lower and upper  $L \times L$ squares, so it is reasonable that more projectors increase the entanglement and this is seen in figure \ref{FIG:VB1}.  That is, more projections moves one closer to the ground state.  Since the ground state clearly has some valence bonds between the two square clusters, unlike the starting valence bond state, one expects more projections means a greater valence bond entropy.
The results are consistent with earlier calculations, the more efficient algorithm allowing access to three more system sizes.  \textcolor{red} {Recall that a valence bond say $(a,b)$ consists of spins at sites $a$ and $b$ where the spin wave function is
\begin{equation}
\frac{1}{\sqrt{2}}(\uparrow_{a} \downarrow_{b} - \downarrow_{b} \uparrow_{a})
\end{equation}
where $\uparrow$ is a spin with z component $\frac{1}{2}$ and $\downarrow$ is a spin with z component  $-\frac{1}{2}$ and such a wave function has total $S_{z} = 0$ and $S=0$.
}

\begin{figure}
\includegraphics[width=0.60\textwidth]{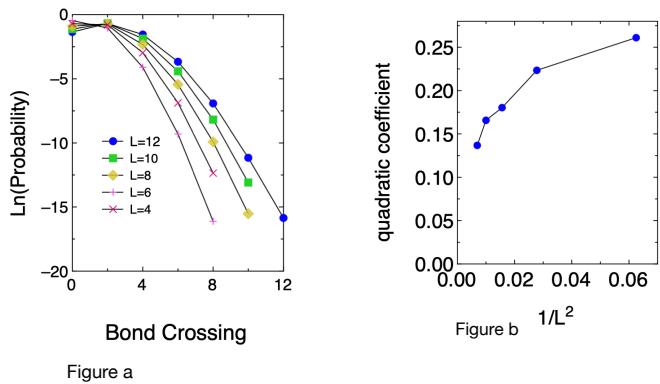}
\caption{  \label{FIG:DIS} Figure a :  distribution of valence bonds for $L \times L-L \times L$ clusters. Blue circles are for   $L=12$, green squares for $L=10$, gold  diamonds $L=8$,  purple x's $L=6$ and pink crosses $L=4$. Figure b : quadratic coefficient of the distribution function  vs. $1/L^2$. The lines in both figures are guides for the eye. }
\end{figure}

To gain further insight, the distribution of connecting bonds was calculated.  In figure \ref{FIG:DIS} {a}, the logarithm of the distribution is plotted for various system sizes.   Taking the logarithm is motivated by intuition provided by the central limit theorem.
Referring to figure \ref{FIG:DIS} {a}, the most probable value of the number of connecting valence bonds is two and the probability of getting two doesn’t change much with system size. (For $L=4$, the most probable value is zero which we take to be a finite size effect).  The feature that changes is the width of the distribution and the curves in figure \ref{FIG:DIS} {a} appear to be quadratic.   Only an even number of bonds are    allowed to cross over from one cluster to the other since all sites must have a valence bond and valence bonds join the A and B sites.  It is also physically plausible that two valence bonds is the most probable value since there is only one interaction joining the two clusters.   Note zero probabilities are not graphed, i.e. for the $6 \times 6 - 6 \times 6$ cluster there are no valence bond configurations with greater than 8 connecting valence bonds (to within the numerical accuracy).  Figure \ref{FIG:DIS} {a} then suggests a distribution of valence bonds crossing from one cluster to the other of the form
\begin{equation}
P(n) = \frac{b}{\sqrt{2 \pi \lambda^2}} \exp{ -\frac{1}{2} (\frac{n-\mu}{\lambda})^{2}} 
\end{equation}
where $n$ is the number of valence bonds that cross from one cluster to the other and $\mu$ is the most probable number of valence bonds.  If $\lambda = c L $,  for large $L$ such a distribution gives the mean number of valence proportional to $L$ in agreement with figure \ref{FIG:VB1}.

Fitting the curves in figure \ref{FIG:DIS} {a}  with a quadratic function,  one finds the coefficient of the quadratic term shows the behavior plotted in figure \ref{FIG:DIS} {b}, where the coefficient is plotted vs $1/L^2$.  Of course, figure \ref{FIG:DIS} {b} is not very linear, this could mean that $12 \times 12 - 12 \times 12$ is not a large enough system size.  However, a line \textcolor{red}{through} the last two points $L = 10$ and $L =12$ tends to point to the origin.  In any case, the mechanism for the anomalous behavior of the valence bond entropy is different between the Bethe cluster and the square cluster.  For every component of the Bethe cluster wave function the number of valence bonds scales as the number of sites in the cluster (see the simple argument in Appendix A0).  On the other hand, for the square clusters, the scaling with $L$ is due to a few components with a very large number of valence bonds, it is a large deviation result from the most probable value.  

\begin{figure}
\includegraphics[width=0.30\textwidth]{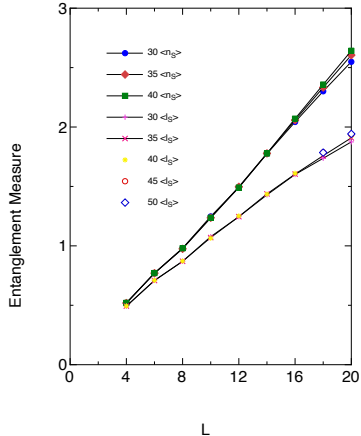}
\caption{ \label{FIG:SQ} Entanglement measure  vs. $L$ for the square cluster.   The blue circles are for calculations of $<n_S>$ with    $2 \times 30$  $L^2$ operators.  The pink plusses are for calculations of $<l_S>$ with  $2 \times 30 $ $L^2$ operators.     The other symbols are explained in the legend.  Lines are guides for the eye. }
\end{figure}

Do other measures of entanglement have similar anomalous behavior?  To address this question in a numerically straightforward way, the quantities defined by Lin and Sandvik \cite{lin} were studied.  In figure \ref{FIG:SQ}, the number of crossing bonds $<n_S>$ for $L \times L - L \times L$ clusters is investigated.  Again loop updates are used with a double projector Monte Carlo so the calculations are more numerically demanding.  Double projector Monte Carlo samples from two independent ground state wave functions making it more time consuming than a single projector, one ground state, method.

 It appears that $<n_S>$ depends linearly on $L$ with reduced values compared to the valence bond entropy. This behavior is consistent with that  observed by Lin and Sandvik for $L \times L$ clusters.  In figure \ref{FIG:SQ} $<l_S>$, the loop entropy \cite{lin}   is  also plotted vs. $L$ and it appears that $<l_S>$ appears much less linear than  $<n_S>$.  This could be interpreted as due to sub leading corrections to a term linear in $L$.  Another possibility is that  $<l_S>$ is not really linear in $L$.  An additional complication is that both  quantities, increase  as the number of projections increase, with $<l_S>$ being more sensitive to the number of projections.  

\section{Bethe Clusters}

What is the behavior of these quantities for the Bethe cluster ?  Before examining these results, recall there are chemical models of the Bethe cluster, i.e. dendrimers \cite{tomalia},\cite{kumar},\cite{chan1}.  In these interesting realizations, very large systems are not possible due to crowding at the boundary \cite{hervet}.  There is also a recent physical model, based on cold atom systems \cite{bentsen}, which, in principle, has no limitation on the size of the cluster.   

\begin{figure}
\includegraphics[width=0.60\textwidth]{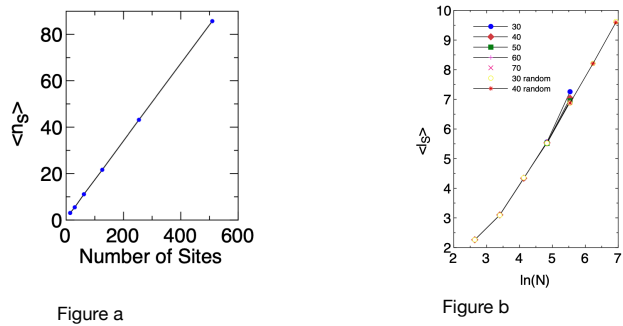}
\caption  { \label{FIG:BS} Figure a : Average number of bond crossing $<n_S>$  vs. \textcolor{red}{number of sites} for the Bethe cluster.  Figure b : Average number of loop crossing  vs. the logarithm of \textcolor{red}{number of sites}  for the Bethe cluster.  The red asterisks  are for projections with a random initial state and $40 \times N$ ($N$ is the system size)  operators , the yellow circles for projections with a random initial state and $30 \times N$ operators.  The other symbols in the legend refer to the entangled initial state and varying number of projections.   The lines are guides for the eye.}
\end{figure}

 In figure \ref{FIG:BS}{a} $<n_S>$ is plotted and one sees linear dependance on the number of sites in the cluster.  The values only depend weakly on the number of projections or the initial states.  This is reasonable since the origin of the anomalous behavior occurs in each of the components of the wave function.  
\textcolor{red} {That is, the ground state wave function, computed by the Monte Carlo procedure, is expanded in a basis of valence bond states.  Since the ground state has spin zero, the basis state used  have valence bonds connecting  the A and B  sub lattices.  By the geometry of the cluster this forces $cN$ valence bonds connecting  the two halves of the Bethe cluster for each such basis state and hence one obtains a value of at least $cN$ for each measurement  of $n_S$  no matter how close one is to the ground state (for a description of the  measurement see appendix A ).}
 
 Note that $<n_S>$ tends to the minimal number of crossing bonds, $N/6$,  $N$ being the number of sites, for large $N$,  that are allowed by the constraint that A and B sites are joined by a bond.  This is consistent with the behavior of the valence bond entropy \cite{friedman}.  \textcolor{red}{The two smallest system sizes, $14$ and $30$ sites are included in the figure to indicate that for $<n_S>$,  linearity starts at small system sizes.}

We now consider  the loop entropy, $<l_S>$ for the Bethe cluster.  In figure \ref{FIG:BS}{b} $<l_S>$ is plotted vs. $ \ln(N)$ where $N$ is the number of sites in the cluster.  The overall message is that $<l_S>$ scales logarithmically  with system size in contrast with the linear behavior of $<n_s>$.  The symbols that occur up to system size 254 refer to projections of 30,40,50,60,70 (more precisely  $N \times 30$. ….).  For system size 254 we see sensitivity to the number of projections,  where somewhat surprisingly, entanglement $\bf{decreases}$ with the number of projections.      

However,  for convenience, the initial state consisted of valence bonds reflected about the central interaction, see figure 1, connecting (A,4 : B,14) (A,3 : B,13) etc.  Since the initial state is very entangled, it is plausible that more projections are needed to attain a more accurate less entangled state.  That is, if the initial state is more entangled than the ground state,  more projections tend to decrease the entanglement since one is moving closer to the ground state.     The additional symbols for system sizes 254, 510, 1022 were 30, 40 projections.  However, here a random initial state consistent with the constraint that valence bonds connect the A and B sites was used.  One sees much less dependance on number of projections and thus larger systems can be accessed.  Again, this is reasonable in that a random state is likely closer in entanglement to the ground state.  For both cases, only a single valence bond state, no superposition, was used as the initial state in the Monte Carlo procedure. 

\begin{figure}
\includegraphics[width=0.30\textwidth]{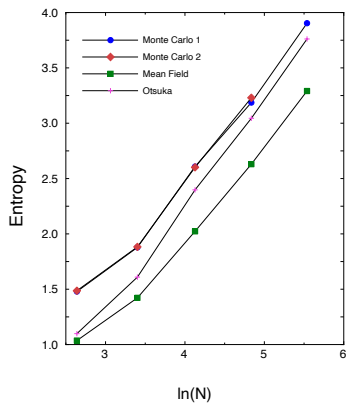}
\caption{\label{FIG:BS2}$S_2$  vs. system size for the Bethe cluster.  The blue circles (Monte Carlo 1) are for  a random initial state and the red diamonds (Monte Carlo 2)  are for the high entanglement initial state.   The green squares are a modified spin-wave calculation of $S_2$ \cite{friedman}. The pink crosses are based on an argument due to Otsuka \cite{otsuka} for the entanglement entropy.  The lines are guides for the eye. }
\end{figure}

\begin{figure}
\includegraphics[width=0.60\textwidth]{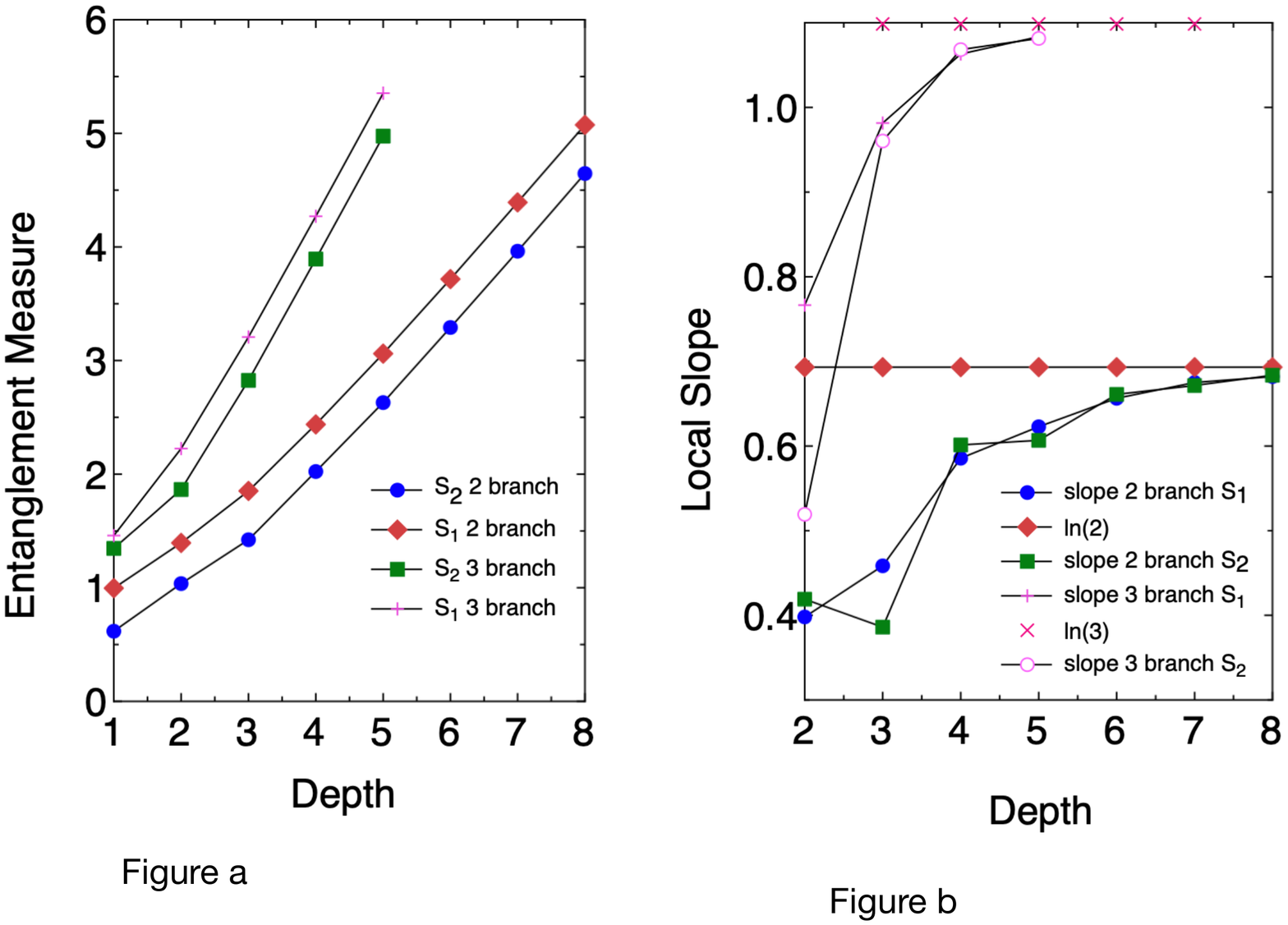}
\caption{\label{FIG:BETMF} Mean field theory for the Bethe Clusters.\\  Figure a is entanglement measure  vs. depth for 2 and  3-branch Bethe clusters.    $S_{2}$, the blue disks,  refers to the second Renyi entropy for the two branch Bethe cluster.  The lines are guides for the eye. \\Figure b is local slope  vs. depth for 2 and  3-branch Bethe clusters.   The blue disks refer to the local slope of the entanglement entropy for the 2 branch Bethe cluster.  The green squares are for the second Renyi entropy for the 2 branch Bethe cluster. \textcolor{red}{All calculations in the figure  were done by modified  spin wave theory \cite{song}.  }   }
\label{fig. 10}
\end{figure}

Due to the above results, the calculations for $S_2$ were reexamined.  An improved algorithm based on loop updates \cite{kallin3},\cite{kallin4} was used and dependance on the initial valence bond state was studied.   In figure \ref{FIG:BS2} $S_2$ is plotted vs $ \ln(N)$ for Bethe clusters.  The calculations using a random initial state and the more entangled initial state are shown and there is little dependence on the initial state for the system sizes shown.  These calculations support the idea that $S_2$ for the Bethe cluster scales as $\ln(N)$, consistent with the loop entropy result and in disagreement with the valence bond entropy.  It should also be noted that modified spin-wave theory \cite{song},\cite{luitz},\cite{takahashi},\cite{hirsch}, a type of mean field theory, gives a $ \ln(N)$ scaling for $S_2$ \cite{friedman} and computational experience \cite{kumar} suggests this should be the case.  In the figure, the mean field theory is plotted as the green squares while $S_2$ is plotted  as blue circles.  For a brief discussion of modified spin wave theory,  see appendix C.

In figure \ref{FIG:BS2}, For $N = 126$ ($\ln(N)$ $\approx 4.84$) $35 \times N$ projections were used with the entangled initial conditions and $30 \times N$ projections were used for the random initial condition.  To within statistical errors, $20 \times N$ projections, for random initial conditions, agreed with the $30 \times N$ results, consistent with the less severe dependance on number of projections shown in figure \ref{FIG:BS}.  We have therefore extended the calculations to $N=254$ sites for $20 \times N$ projections and random initial conditions. Our previous calculations (see figure 8 of  \cite{friedman} ) with the entangled initial conditions and  $20 \times N$ projections gave a substantial larger value of $S_2$ , i.e. 4.9 with a small statistical error.  Furthermore, the next larger system size $N= 510$ gave $S_2$ = 7.4 $\pm .4$.  The conclusion we made, now which we believe  to be wrong, was that  $S_2$ scales as $N$ and was based on the last two system sizes where an insufficient number of projections were used.  

There is another interesting argument due to Otsuka \cite{otsuka} that favors $\ln(N)$ behavior.  If the Bethe cluster is bisected, for example, between (B,8) and (A,7) in figure 1, it is then straightforward to see that absolute value of the number of A sites minus the number of B sites $| N_A - N_B |$ is 
\begin{equation}
|N_A - N_B| =   |\sum_{n=0}^{l}(-1)^{n}2^{n}| = |\frac{1-(-2)^{l+1}}{3}|
\end {equation}.   
Here $l$ is the depth of the cluster ,  i.e. 2 in figure 1.   \textcolor{red} {However, for such  a bond centered Bethe cluster,  the total number of A and B sites,  are equal \cite{Changlani} .This local mismatch between the number of A and B sites leads to dangling bonds on the boundary, whose presence increases the number of quasi degenerate ground states \cite{Henley}.   Evidence is presented in \cite{Changlani}, \cite{Henley}, that these quasi degenerate states lead to a singlet triplet gap that scales as $1/N^{2}$, $N$ being the number of sites in the cluster, in contrast to the $1/N$ scaling for a square cluster.   }

 If one assumes a Lieb Mattis theorem for the highest weight state of the reduced density matrix then the highest weight (largest eigenvalue) state has spin $
S = | N_A - N_B |$/2.  We are unaware of a rigorous justification, however, it is intuitively appealing  and numerical evidence suggests it is true \cite{otsuka} .  Hence for large $l$, the highest weight state has degeneracy $2S+1$ approaching $2^{l+1}/3 $.  It is therefore plausible that the minimum number of states to approximate the ground state (for  half the cluster) is at least      $2^{l+1}/3 $.  Asserting a Boltzmann like formula for the entanglement entropy $ S_1 \approx \ln(n_{approx})$ one sees $S_1 \ge \ln(2^{l+1})$  i.e. $S_1 \ge l\ln2$.  Hence the entanglement entropy should scale at least linearly with the depth.  In figure 9, the pink crosses is $\ln(2S+1) = \ln(|\frac{1-(-2)^{l+1}}{3}|+1)$.  It appears there is a rather close correspondence to the Monte Carlo calculation of $S_2$.   Due to Lieb's theorem \cite{lieb2}, the same argument implies for a repulsive Hubbard model at $1/2$ filling on the Bethe cluster,  $S_1$ scales at least as the depth of the cluster. 

The Otsuka type argument can also be applied to a 3 branched cluster yielding an entanglement entropy that scales as $S_1 \approx l \ln(3)$.  This can be checked with a modified spin-wave calculation as shown in figure \ref{FIG:BETMF} {a} where entanglement entropy vs. depth of the cluster is plotted for 2 and 3  branched Bethe clusters.  There is linear behavior in both cases with an obvious  larger slope for three branches.   To make this clearer, in figure \ref{FIG:BETMF} {b} the local slope ( $S_1(l+1)-S_1(l)$) is plotted.  One sees that the local slope  approaches $\ln 2$ for 2 branches and $\ln 3$ for 3 branches.

Numerically, this logarithmic scaling is more  restrictive for practical computability  than the $\ln l$ scaling for say a critical linear chain.   That is, a linear chain of length 100 sites  needs $\exp^{\ln(100)} = 100$ states to accurately approximate the ground state while a Bethe cluster of depth 100 needs $e^{100}$ states. 

\section{Return to the $L \times L$ clusters, modified spin-wave theory}

\begin{figure}
\includegraphics[width=0.80\textwidth]{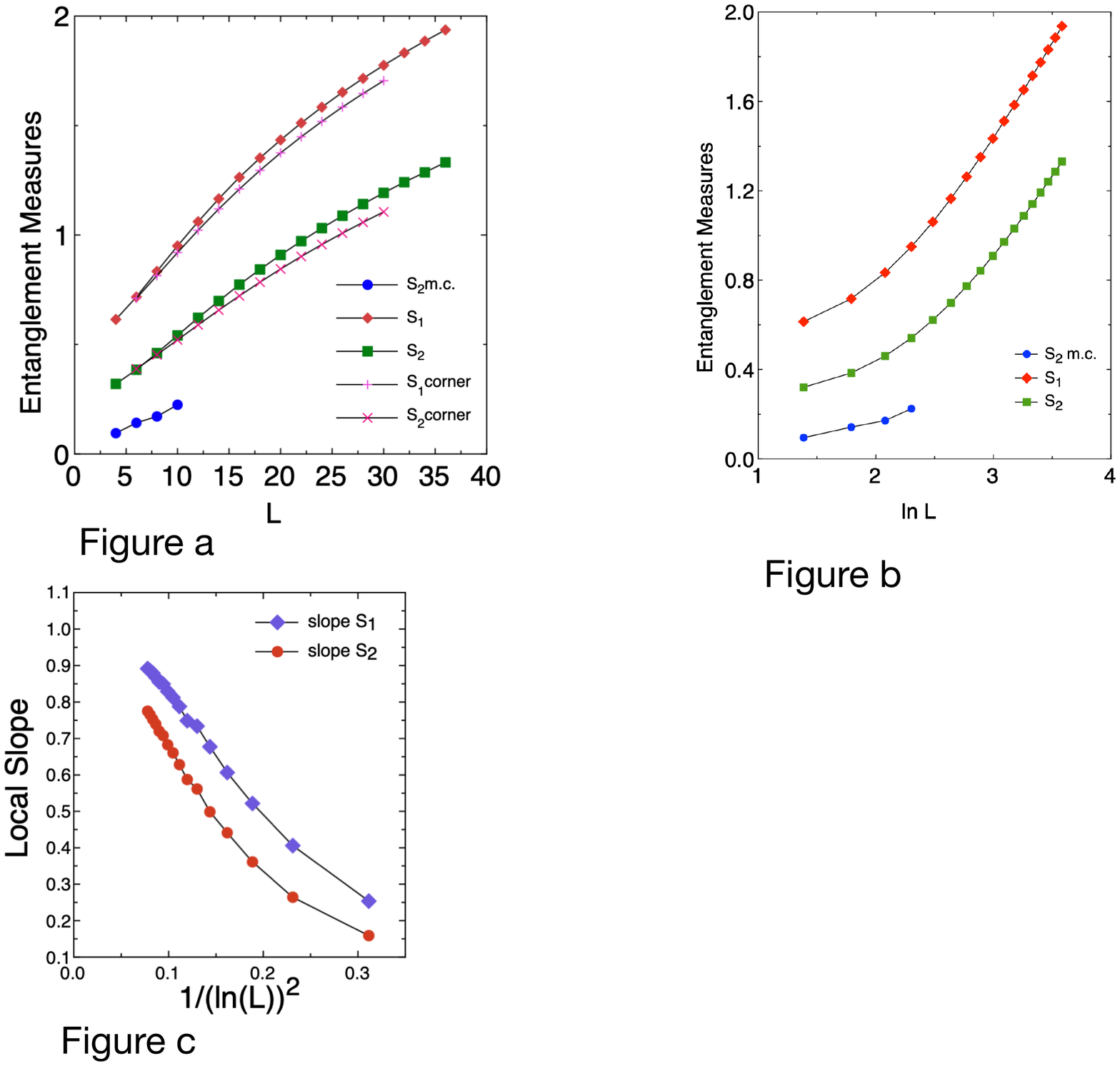}
\caption{\label{FIG:HEIS}Heisenberg model for the $L \times L - L \times L$ cluster.  \\ Figure a : Entanglement measures  vs. $L$ .   The red diamonds are a modified spin-wave calculation for the entanglement entropy $S_1$ while the green squares are a modified spin-wave calculation for $S_2$.   The pink X s and  crosses   are $S_2$ ($S_1$) with the single connection occurring between the corners of the squares as opposed to mid points.The blue circles are Monte Carlo results for $S_2$ with statistical errors the order of the size of the symbols.  \\
Figure b :  Entanglement measures  vs. $\ln L$.   The red diamonds are a modified spin-wave calculation for the entanglement entropy $S_1$ while the green squares are a modified spin-wave calculation for $S_2$.  The blue disks are Monte Carlo results for $S_2$ with statistical errors the order of the size of the symbols. \\
Figure c : Local slope  vs. $\frac{1}{(\ln L)^2}$.   The lines are guides for the eye.   }
\label{fig. 12}
\end{figure}

 Finally we return to the $L \times L - L \times L$ clusters.  In figure \ref{FIG:HEIS} {a}
 $S_2$,  blue disks,  is plotted vs $L$ for the $L \times L - L \times L$ clusters.  A low entropy state is used as the starting state.  One sees  behavior   consistent  with linear  $L$ in agreement with our previous Monte Carlo calculation.  However, the number of systems sizes considered and the largest system size ( $L$ = 10) is limited.  To explore larger system sizes, mean field theory with system sizes up to 36 are explored in the same figure, the green squares being $S_2$ and the red diamonds being $S_1$, the entanglement entropy.  The pink X s and  crosses   are $S_2$ ($S_1$) with the single connection occurring between the corners of the squares as opposed to mid points.  As far as the scaling is concerned, there doesn't seem much difference where the connection is made.  \textcolor{red}{Note that the corners are connected by two bonds to their respective clusters while the mid - points  are connected by three bonds.  We suggest this  number of connections, for sufficiently large clusters, determines the constant term in the entanglement entropy.  A natural to conjecture is that more connections gives more entanglement and hence a bigger constant.}
 
\begin{figure}
\includegraphics[width=0.80\textwidth]{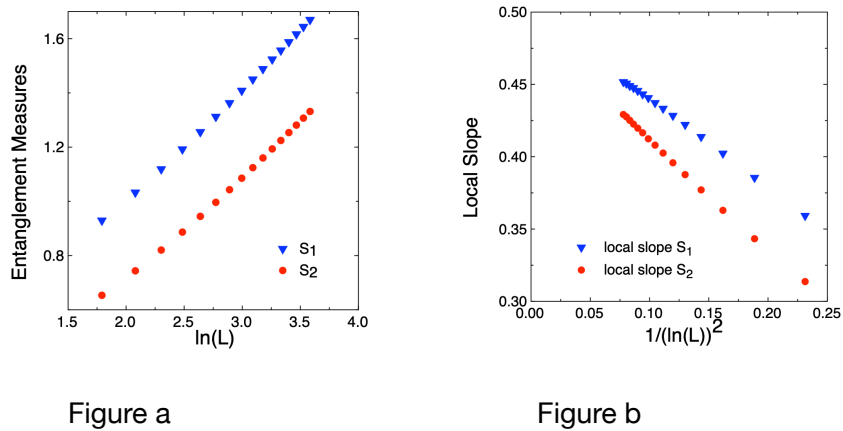}
\caption{\label{FIG:XY}$X-Y$ model for the $L \times L - L \times L$ cluster.  Figure a : $S_1$ and $S_2$  vs. $\ln L$.    The blue triangles are a modified spin-wave calculation for the entanglement entropy $S_1$ while the red circles are a modified spin-wave calculation for $S_2$. Figure b : Local slope  vs. $\frac{1}{(\ln L)^2}$.    }
\end{figure}

Most significantly,  one sees a downward curvature in the plot calling the linear scaling into question.  To explore this issue, the same calculations are plotted vs. $\ln(L)$ in figure \ref{FIG:HEIS} ${\bf b}$.  In this figure, there appears an upward curvature, in fact, if one plots vs. $(\ln(L))^2$ the graph appears more linear.  However, we will argue that the scaling is actually a pure logarithm with noticeable finite size effects.   In figure \ref{FIG:HEIS} ${\bf c}$ , the local slope vs. $\frac{1}{(\ln L)^2}$
is plotted.  The local slope for the entanglement entropy is defined to be $\frac{S_1(L)-S_1(L-1)}{\ln(L)-\ln(L-1)}$ with an analogous definition for the second Renyi entropy. As $L$ becomes large the local slope appears to approach range  between $1$ and $1.1$.  
\textcolor{red} {To obtain a more precise value, an extrapolation has been done  (a linear least squares fit) for system sizes 24 - 36.
The constant, in the fit, is the extrapolation for $L$ going to infinity, and we find 
$S_1 = 1.11 \pm .01$
$S_2 = 1.11 \pm .01$.}

Larger system sizes, hard to access numerically even for a modified spin wave calculation, are needed to reach a definitive conclusion from numerical evidence.  However, there are theoretical reasons to prefer $1$, that is, $S_1  \approx \frac{n_G}{2} \ln{L}$  where $n_G$ is the number of Goldstone modes and $n_G = 2$ in the case of two dimensional Heisenberg anti ferromagnet.     

Analytical and numerical arguments\textcolor{red}{\cite{Met}},\cite{luitz} give that the leading corrections to the area law term for a subsystem $A$ of perimeter $l_A$  embedded in a much larger cluster, goes as  $\frac{n_G}{2} \ln{l_A}$ .    We suggest that in our case, the leading term goes to zero since $l_A$ is zero but the sub leading term is retained with $l_A$ replaced by $L$.   This is natural since for 2 $L \times L$ clusters joined by $L$ bonds there is a  $\frac{n_G}{2} \ln{L}$ and an $a L $ term.  From a single connection, the $a L$ term can grow one link at a time, however there is no obvious way a $\ln(L)$ can grow by adding links, hence it must be present even for one link. 

To verify  these considerations, another model on $L \times L - L \times L$ clusters was considered, namely, the ferromagnetic $X-Y$ model. 
\textcolor{red}{In particular, the proportionality on the number of Goldstone modes was checked.}
The ground state  of this model is unique for finite clusters \cite{mattis} with the Hamiltonian given by

\begin{equation}
H =  J \sum_{<i,j>} \bf{S_i^ X} \bf{S_j^X} +  \bf{S_i^Y} \bf{S_j^Y}
\end {equation} 
where  $<i,j>$ refer to nearest neighbors
with  $J  <  0$  and the spin operators have spin $1/2$. 

The ground state entanglement properties  can be treated with large scale Monte Carlo methods \cite{bohdan} and modified spin wave theory \cite{luitz} .  (For spin-wave theory of the ground state of the $ X-Y$ model see \cite{gomez}).  From the experience with the Heisenberg model, the modified spin wave theory was used, since to approach the asymptotic limit relatively large system sizes are necessary.  In figure \ref{FIG:XY} ${\bf a}$ ,  $S_1$ and $S_2$ are plotted vs.  $\ln(L)$.  As seen from the figure, there is an upward curvature even for the largest system sizes i.e. $L=36$.  Apart from dealing with large matrices, the small value of the field needed to restore $U(1)$ symmetry \cite{luitz}, makes larger  calculations difficult.   The special choice of the cluster geometries in \cite{luitz}, \cite{bauer}  permit a  more analytical treatment; no matrix diagonalizations are needed which makes arbitrary precision numerics feasible.  Note that the subsystem in   \cite{luitz}, \cite{bauer}  is a line, see \cite{berthiere} for interesting subtleties for this subsystem.

To address the curvature issue, the local slope is plotted in figure \ref{FIG:XY} ${\bf b}$ .  As $L$ becomes large, the local slope  for both $S_1$ and $S_2$ appears to approach $1/2$.  \textcolor{red} {A linear least squares fit gives $S_1 = .493 \pm .001,  
S_2 = .492 \pm .001$}. This is again consistent with theoretical considerations, for $U(1)$ symmetry, the symmetry of the $X-Y$ model, the number of Goldstone models $n_G$ is one.

\section{Conclusions}

We have studied  the ground state  entanglement  properties, for the spin $1/2$ antiferromagnetic Heisenberg model, on Bethe clusters and $L \times L  -  L \times L$ clusters.  The computational approach  was  an improved quantum Monte Carlo method \cite{evertz}, \cite{kallin4} and modified spin wave theory \cite{song}, \cite{luitz}.     The entanglement measures introduced by Lin and Sandvik \cite{lin}, in addition to the valence bond entropy and $S_2$ , the second Renyi entropy, were studied.
For the bisecting of the Bethe cluster, in disagreement with our previous results \cite{friedman} and in agreement with modified spin-wave theory and Otsuka's argument \cite{otsuka} , the valence loop entropy and the second Renyi entropy is proportional to  the logarithm of the number of sites in the cluster.  By using modified spin-wave theory, the coefficient of the logarithm was shown to approach $\ln 2$ and $\ln 3$ for the two and three branched Bethe Clusters.  Recall that the essential idea of modified spin-wave theory \cite{song},\cite{luitz}   is to introduce a staggered field to  approximately maintain spin rotational invariance, for finite clusters,   which is  otherwise broken   by  a spin-wave approach.

That $S_2$ should scale as $ \ln N$, $N$ being the system size, \textcolor{red} {is suggested by on the basis of computational studies} (for example see \cite{ kumar} ) where dmrg appears to be quite accurate.  If say $S_2$ scaled as the system size such accuracy by dmrg  should be unattainable.  One of the authors (B. F.)   previous argument for scaling with $N$ was based on quantum Monte Carlo with an insufficient number of projections for an initial state having high entanglement.
   
 The results for the $L \times L  -  L \times L$ clusters are in agreement with our previous results, a linear scaling of the valence bond  entropy with $L$ where the mechanism is rare states with a large number of valence bonds.  However, a more refined measure of entanglement, the valence loop entropy indicates a slower growth  with $L$, and the system sizes accessible to $S_2$ are too small to distinguish scaling with $L$ from weaker size dependance.  To access  larger system sizes, modified spin-wave theory  was again used.  These calculations again indicate a weaker scaling with $L$ , namely,  $S_1 (S_2)  \approx \frac{n_G}{2} \ln{L}$  where $n_G$ is the number of Goldstone modes, $n_G = 2$ for the antiferromagnetic Heisenberg model.  The coefficient of the logarithm is universal and is the analogue of the sub leading term for bisecting  a $L \times L$ cluster \textcolor{red}{\cite{Met}}, \cite{luitz}.  To further verify this result, the ferromagnetic $X-Y$ model was investigated, giving again  $S_1 (S_2) \approx \frac{n_G}{2} \ln{L}$. In this case the calculations indicate $n_G$ equals one, the number of Goldstone modes for the $U(1)$ symmetry of the $X-Y$ model.  \textcolor {red} {A summary of our calculations  is given in Table 1.}
  
Taken together, the results for the Bethe clusters and the $L \times L  - L \times L$ clusters indicate that the scaling of the valence bond entropy can be quite different from the entanglement entropy, i.e. linear dependence can be replaced by logarithms.  
The calculations suggest linking high entanglement objects will not generate much more  entanglement.  As a consequence, simulating the ground state of clusters joined together by a few bonds should not be much more difficult than simulating the ground state of a single cluster.  In quantum chemistry, a recent review of theoretical and computational evidence  \cite{chan} , suggests no exponential quantum speed up for the ground state properties of large molecules.  If one views large molecules as clusters linked together by a few bonds, our results are consistent with the studies from quantum chemistry.

\section{acknowledgments}

We thank Dr. M. Stoudenmire and Professor N. Laflorencie  for very helpful correspondence.

\section{Appendix A0}
A few years ago, it was noticed numerically, that for clusters of non interacting fermions, at half filling,  joined by a single bond \cite{levine}, the entanglement entropy scales as the number of sites  for the Bethe cluster (away from half filling see \cite{schreiber}) and the square root of the number of sites, $L$ for a $L \times L - L \times L$ cluster  \cite{caravan}.  These results can be understood by an argument involving zero energy single particle states.  For example, in figure \ref{FIG:CLUSTER} ${\bf a}$ there is a zero energy single particle state $ \phi_1$ with amplitude $\frac{1}{\sqrt{2}}$ on site (A,1), amplitude $-\frac{1}{\sqrt{2}}$ on (A,2) and zero on any other site.  There is also an analogous state $\phi_2$ on site (B,11) and (B,12).  A numerical calculation gives  states  which are superpositions of the zero energy states and these states are occupied at half filling.  This leads to entanglement between the top and bottom half of the cluster that scales as $N$,  the number of sites in the cluster.

 In addition, for an isotropic quantum Heisenberg model on a balanced Bethe cluster, the valence bond entropy scales as the number of sites in the cluster \cite{caravan}.  The later result can be seen by a simple argument.  For the ground state wave function written in the valence bond basis,  one can sample the components with the numerical weights of the component, which can chosen to be positive. The average number of valence bonds connecting the two halves of the cluster (see figure \ref{FIG:CLUSTER} ${\bf a}$ for the relevant Bethe cluster, and figure \ref{FIG:CLUSTER} ${\bf b}$ for the square clusters) is the valence bond entropy divided by $ \ln 2 $. A bipartite cluster with equal number of A and B sub lattice sites is called a balanced cluster and the ground state of an isotropic quantum Heisenberg model on a balanced cluster has spin zero \cite{lieb}.  For eigenstates with spin zero, for each basis state, the valence bonds  join the A and B sub lattices.
 
 This forces $ c N$
valence bonds (where $c$ is a constant which tends to $1/6$  for large clusters)  to connect the two halves since there are unbalanced numbers of A and B sites between the two halves of the cluster.  Of course, this argument only works for the Bethe cluster.  Numerically, it appears that even for two square  $L \times L$  clusters, the valence bond entropy scales as $L$ though this is not forced by the geometry of the cluster \cite{friedman}.  In both examples, the valence bond entropy scales as the sites in the boundary, not the sites in the boundary between the clusters.  Recall for the Bethe cluster, there are many boundary points, i.e. the number of boundary points scales as the number of sites in the cluster.

\section{Appendix A}
 \textcolor {red} {This appendix is a brief description  of the loop algorithm and how observables are calculated from the loop algorithm}.  The quantum Monte Carlo method  starts from a sufficiently high power of $H_{proj}$ applied to a valence bond state, that is,   $H_{proj}^{m} |vb>$ where 

\begin{equation}
H_{proj} = \sum_{<i,j>} ({\bf{S_ i} \cdot \bf{S_j}}  - 1/4)
 =-\sum_{<i,j>} H_{ij}
\end{equation}

In the loop algorithm \cite{evertz}, unlike Sandvik's original approach \cite{sandvik} $H_{ab}$ is split into diagonal and off diagonal pieces

\begin{equation}
H_{ab}(1)= 1/4 - {\bf S_{a}^{z}S_{b}^{z}}
\end{equation}
\begin{equation}
H_{ab}(2)=-1/2 \  ({\bf S_{a}^{+}S_{b}^{-} + S_{b}^{+}S_{a}^{-}})
\end{equation}

The z component of spin, as well  as the valence bonds, are considered in the basis states, i.e. a mixed basis is used.  The sums in $H_{proj}^{m} |vb>$ are then expanded so a product of sums becomes a sum of products

\begin{equation}
H_{proj}^{m} |vb> = \sum_{r=1}^{N_{nn}^{m}} P_{r} |vb>
\end{equation} 

\begin{figure}
\includegraphics[width=0.40\textwidth]{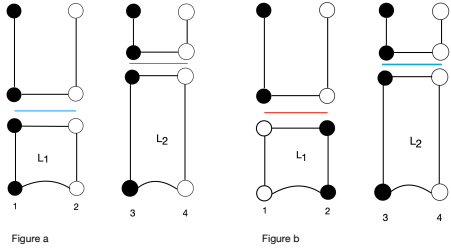}
\caption{ \label{FIG:LOOP} 
Loop diagrams.  Figure a is the initial loop diagram.  Figure b is loop diagram after the spin flip.
}
\end{figure}

The m operators in $P_{r}$ are either  $H_{i_k j_k}(1)$ or $H_{i_k j_k}(2)$  $k =1, ......m$ and $N_{nn}$ is the number of nearest neighbor sites.  The task is then to sample the sum.  An efficient way to do this is illustrated in figure \ref{FIG:LOOP} for a 4 site system  m=2 and the observable of interest is the valence bond entropy.  In particular, $ |vb> = |(12)(34)>$ and $P_{r}$ is $H_{34}(1)H_{12}(1)$ for figure \ref{FIG:LOOP} {\bf{a}} .  

In the diagram, filled (empty) circles represent a value of  $S_z = 1/2 (-1/2)$, the bottom state is   $|(12)(34)>$ and the top state is the Neel state.  The blue (red) lines represent an interaction of type 1 (2).  To sample the sum the loops are flipped with probability 1/2, all spins in the loop change from $S_z$ to $-S_z$ and therefore interactions change from type 1 (2) to type 2 (1). For example, if loop $L_1$ is flipped (but $L_2$ is not) the second interaction changes from blue to red in the figure.  After the loops are flipped, loop updates, diagonal interactions are updated randomly and a new diagram is constructed.  The method is motivated by the stochastic series expansion (SSE).  For a detailed justification see \cite{evertz}  and further explanation see \cite{kallin2}.  We have checked this procedure against the more logically straightforward Sandvik algorithm and the two methods are consistent with the loop algorithm being numerically more efficient \cite{kallin2}.  The procedure described above corresponds to single projector quantum Monte Carlo, for double projector quantum Monte Carlo, the top state, the Neel state, is replaced by a valence bond state.  The top state is then included in loops and therefore flips are permitted.  

\textcolor{red} {How do you make measurements from such diagrams ? Let us start with the simplest possible observable, the valence bond entropy.  In this case, one uses a diagram like figure \ref{FIG:LOOP} {\bf{a}} or $\bf{b}$, the top state is the Neel state.  The diagram in figure \ref{FIG:LOOP} {\bf{b}} corresponds to the product of operators $H_{34}(1) H_{12}(2)$.  In a more realistic example, after one sweep (a combination of loop updates and random updates for diagonal terms ), one gets a product of $2n$ operators,  (for $2n$ projections) }
\textcolor{red}{
\begin{equation}
H_{i_{2n},j_{2n}} (\lambda_{2n}) \cdots H_{i_{2},j_{2}} (\lambda_{2})H_{i_{1},j_{1}} (\lambda_{1})
\end{equation}  
}

\textcolor{red} {
Here the $\lambda$'s take the value 1 or 2 depending on whether the operators correspond to a blue or red line.  To make a measurement one replaces each $H_{ab}(\lambda)$ by the full $H_{ab}$ and propagates the initial valence bond state by $n$ operators ($1/2$ the total number of operators).  A new valence bond state is obtained and for  the new   valence bond state one counts the number of bonds crossing from the subsystem to the environment.}

\textcolor{red} {This quantity is then averaged over all sweeps.  In the case of $<n_S>$, the measurement is the same, however, the loop updates are different.  The diagram in figure \ref{FIG:LOOP} a is capped with a valence bond state and loops containing this valence bond state  are allowed to be flipped.  To calculate the valence loop entropy  $<l_S>$, one propagates the valence bond state from the top and the valence bond state from the bottom and then calculates, in the middle, the overlap diagram as described in  Appendix B.  The loop entropy is the average of the number of loops  crossing from the subsystem to the environment in the overlap diagram.   
To calculate the second Renyi entropy $S_2$ we used loop updates and the ratio technique  \cite{kallin2},\cite{kallin3},\cite{kallin4}.  Recall 
\begin{equation}
S_{2}(\rho_{A}) = -\ln[Tr \rho_{A}^{2}]
\end{equation} 
}

\textcolor{red} {
where $\rho_{A}$ is the reduced density matrix for the subsystem $A$.  By introducing two copies of the system one finds
\begin{equation}
 -\ln[Tr \rho_{A}^{2}] = -\ln (<Swap_{A}>)
\end{equation} 
where $Swap_A $ interchanges endpoints of valence bonds within $A$ between the copies. }

\textcolor{red} { The ratio technique is a method where one starts from  smaller subsystems, which are easier to calculate, and then proceed to larger subsystems by a sequence of ratios. Say $A_{m-1} \subset  A_{m}$ and then one calculates  $\frac{<Swap_{A_{m}}>}{<Swap_{A_{m-1}}>}$ which allows one to obtain $<Swap_{A_{m}}>$.
That is, 
\begin{eqnarray}
S_{2}(\rho_{A_{m}}) = -\ln<Swap_{A_{m}}>  =                            \\
-\ln<Swap_{A_{1}}>  -\ln \frac{<Swap_{A_{2}}>}{<Swap_{A_{1}}>}  \cdot \cdot \cdot   \nonumber\\
-\ln \frac{<Swap_{A_{m}}>}{<Swap_{A_{m-1}}>}  \nonumber
\end{eqnarray} }

\textcolor{red} {The ratio$\frac{<Swap_{A_{m}}>}{<Swap_{A_{m-1}}>}$ is calculated using a diagram (\ref{FIG:LOOP} {\bf{a-b}}, figure 1 \cite{kallin4}) where a swap operator for $A_{m-1}$ is inserted in the middle of the diagram (see figure 2 \cite{kallin4}).  (Note in \cite{kallin4} the valence bond states are placed on the left and right, the diagram is horizontal as opposed to vertical. )  $\frac{<Swap_{A_{m}}>}{<Swap_{A_{m-1}}>}$ is obtained by propagating the valence bond states  from the top and bottom , meeting in the middle where where an operator is measured which swaps sites in $A_{m}$ which have not already been swapped in $A_{m-1}$.   }

\section{Appendix B}

\begin{figure}
\includegraphics[width=0.30\textwidth]{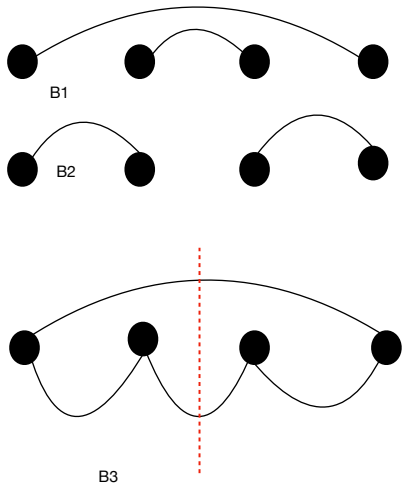}
\caption{overlap diagram}
\label{fig. 19}
\end{figure}

We briefly discuss the loop entropy \cite{lin}.  The loop quantum Monte Carlo algorithm generates two valence bond states, call them $|v_{1}>$ , $|v_{2}>$.   The overlap diagram, discussed below, is then constructed and the number of loops crossing from the subsystem to the environment is counted.  The number of loops  is then averaged over the valence bond states generated by the loop algorithm. To avoid confusion, we emphasize that the loop algorithm is distinct from the loop entropy, i.e. Sandvik's original    method \cite{sandvik} can be used to compute the loop entropy.  To illustrate the loop entropy, consider a one dimension 4 site system.  The subsystem consists of sites of $1$ and $2$ while the environment is sites $3$ and $4$.   Take for example $|v_{1}>$ = $|(14)(23)>$ , $|v_{2}>$ = $|(12)(34)>$, see the valence bond diagrams B1, B2.  Using B1 and B2 the overlap diagram B3 is formed.  We see that there is one loop linking the subsystem to the environment.  If we form the overlap diagram of $|v_{1}>$ with  itself there are two such loops,  while  for $|v_{2}>$ with  itself there are no loops.  

The loop entropy was introduced as an easier way to compute a quantity that has a greater correspondence to the second Renyi entropy and the entanglement  entropy.  In particular, numerical calculations show that if one bisects a $L \times L$ isotropic Heisenberg model the valence bond entropy scales as $L \ln{L}$  \cite{alet,chhajlany} as opposed to $S_2$ which scales as $L$ \cite{kallin5}.   However, $<l_S> $ has no such extra $\ln L $ factor and its numerical value is closer to $S_2$ \cite{lin}.  

\section{Appendix C}

This appendix is a brief review of modified spin wave theory.  For a detailed discussion see \cite{song}.  Starting from the Heisenberg Hamiltonian (1), spin operators $ \bf{S_ r}$ are replaced by boson creation and annihilation operators $b_r$, $b_{r}^{+}$ via a Dyson-Maleev transformation  giving
\begin{equation}
H_{LSW} = E_{Neel} +  \sum_{i,j} [ A_{ij} b_{i}^{+}b_{j} + \frac{1}{2} B_{ij}(b_{i}b_{j } + h.c.)  ]
\end {equation} 
where quartic terms in the boson operators have been neglected (eq. (26) of \cite{song}).  Here $E_{Neel}$ is a constant, $A_{ij} = (h + S \sum_{k} J_{ik}) \delta_{ij}$, $B_{ij} = -S J_{ij}$.  A staggered field $h$ has been added to (1), i.e. a term $h \sum_{i} (-1)^{i}  \bf{S_{i}^{z}}$.  Since $H_{LSW}$ is a quadratic form, it can be diagonalized and the ground state determined.  For the clusters considered, this needs to be done numerically and the staggered field is chosen to restore sub lattice symmetry in an average sense.  To calculate the Renyi entropies, the matrix $C_{rr'}$ is introduced where $r$ and $r'$   are sites in the subsystem $\Omega$.  Here 
\begin{equation}
C_{rr'} = \sum_{r'' \in  \Omega} X_{r r''} P_{r'' r'}
\end{equation}
with $ X_{r r'} =  <(b_{r} + b_{r}^+)(b_{r'} + b_{r'}^+)> /2$ and $P_{r r'} =  - <(b_{r} - b_{r}^+)(b_{r'} - b_{r'}^+)> /2 $. with < > denoting a ground state expectation value.
The Renyi entropies are determined by the eigenvalues $\nu_{q}$ of $C_{rr'}$ \cite{peschel} and  in particular, $S_{2} = \sum_{q} \ln{2\nu_q}$.  For $L \times L   -  L \times  L$ clusters we observe numerically only two eigenvalues grow with $L$ and the other eigenvalues take values close to $\frac{1}{2}$.  Assuming the large eigenvalues scale as $L^{\frac{1}{2}}$ \cite{luitz} gives $S_{2} \approx \frac{n_G}{2} \ln{L}$.






\newpage

\textcolor{red}{Table 1.  Summary of results for entanglement measures.  A question mark denotes a quantity that wasn't calculated. Constants (i.e. $c_1$) with a subscript indicate quantities which weren't precisely determined.}
\\*[1cm]

\begin{tabular}{r|c|c|c|c}  \hline
 & $L \times L  - L \times L$  &  $L \times L  - L \times L$  & Bethe \   2-leg \ depth(n) & Bethe  3-leg \ depth(n) \\ \hline
 &Heisenberg & X-Y & Heisenberg & Heisenberg \\  \hline
 
$S_2$ & $\ln L$ & $1/2\ln L$ &$ (\ln{2}) n$ & $ (\ln{3}) n$ \\ \hline
$S_1$ & $\ln L$ & $1/2\ln L$  & $ (\ln{2}) n$  & $  (\ln{3}) n$ \\ \hline
$S_{VB}$ & $c_{1}L$ & ? & $2^{n+1}/3$ & ? \\  \hline
$<n_{S}>$ & $c_{2}L$ & ? &   $2^{n+1}/3$ & ?\\ \hline
$<l_{S}>$ &   < $c_{3}L$ & ? &  $c_{4} n$ & ? \\  \hline
\end{tabular} 


\begin{references}

\bibitem{otsuka}
H. Otsuka, Density-matrix renormalization-group study of the spin-1/2 XXZ anti ferromagnet on the Bethe lattice, Phys. Rev. B 53, 14004, (1996).


\bibitem{friedman1}
B. Friedman, A density matrix renormalization group approach to interacting quantum systems on Cayley trees, J. Phys. Condens. Matter 9, 9021 (1997).

 \bibitem{kumar} 
 M. Kumar, S. Ramasesha, and Z. G. Soos, Density matrix renormalization group algorithm for Bethe lattices of spin-1/2 or spin-1 sites with Heisenberg antiferromagnetic exchange,  
Phys. Rev. B 85, 134415, (2012).  

 \bibitem{chan1} N. Nakatani and G. K.-L. Chan, Efficient tree tensor network states (TTNS)  for quantum chemistry : Generalizations of the density matrix renormalization group algorithm,  J. Chem. Phys. 138, 134113 (2013).
 
  \bibitem{levine}
  G. C. Levine, and D. J. Miller, Zero-dimensional area law in a gapless Fermionic system, 
  Phys. Rev. B 77, 205119, (2008).


\bibitem{caravan}
  B. Caravan, B. A. Friedman, and G. C. Levine, Scaling of entanglement entropy in point-contact, free-fermion systems,
  Phys. Rev. A 89, 052305, (2014).
 
  
   \bibitem{eisert}
  J. Eisert, M. Cramer, and M. B. Plenio, Colloquium : Area laws for the entanglement entropy,
  Rev. Mod. Phys. 82, 277, (2010).


  

  
  



\bibitem{hastings1}
M. B. Hastings, An Area Law for One Dimensional Quantum Systems, J. Stat. Mech. : Theo. Exp. (2007) P08024.


\bibitem{arad}
I. Arad, Z. Landau, and U. Vazirani, Improved one-dimensional area law for frustration-free systems,  Phys. Rev. B 85, 195145 (2012). 


\bibitem{movassagh}
R. Movassagh and P. W. Shor, Supercritical entanglement in local systems: Counterexample to the area law for quantum matter,  Proc. Natl. Acad. Sci. USA 113, 13278 (2016).
\textcolor{red}{
\bibitem{ramirez} G. Ramirez, J. Rodriguez-Laguna, and G. Sierra, Entanglement over the rainbow, Journal of Statistical Mechanics : Theory and Experiment, Vol. 6,  article id 06002 (2015), arXiv :1502.02695.}




  

 \bibitem{sandvik}
  A. W. Sandvik,  Ground State Projection of Quantum Spin Systems in the Valence-Bond Basis, Phys. Rev. Lett. 95, 207203, (2005).
 
  

  
  \bibitem{hastings2}
  M. B. Hastings, I. Gonzalez, A. B. Kallin, and R. G. Melko, Measuring Renyi Entropy in Quantum Monte Carlo Simulations, 
  Phys. Rev. Lett. 104, 157201, (2010).

  
   
  \bibitem{kallin2}
  A. B. Kallin, Measuring Entanglement Entropy in Valence Bond Quantum Monte Carlo simulations, Masters Thesis, Waterloo University, 2010.   
  
  \bibitem{kallin3}
  A. B. Kallin, Methods for the Measurement of Entanglement in Condensed Matter Systems, Ph.D. Thesis, Waterloo University, 2014.  
  
   \bibitem{friedman}
  B. A. Friedman and G. C. Levine, Scaling of Entanglement Entropy for the Heisenberg Model on Clusters Joined by Point Contacts: 
Possible Violation of the Area Law in Dimensions Greater than One,  Journal of Stat. Phys. 165, 727 -739 (2016).
  
    \bibitem{lin} 
  Y.-C. Lin and A. W. Sandvik, Definitions of entanglement entropy of spin systems in the valence-bond basis, Phys. Rev. B 82, 224414 (2010).
  
  \bibitem{evertz}
  A. W. Sandvik and H. G. Evertz, Loop updates for variational and projector Monte Carlo simulations in the valence-bond basis, Phys. Rev. B 82, 024407 (2010).
  

  \bibitem{kallin4} 
A. B. Kallin, M. B. Hastings, R. G. Melko, R. R. P. Singh, Anomalies in the entanglement properties of the square-lattice Heisenberg model,  Phys. Rev. B 84 16, 165134 (2011).

 \bibitem{white} S. R. White, Density-matrix algorithms for quantum renormalization groups, Phys. Rev. B 48, 10345 (1993); U. Schollwock,  The density-matrix renormalization group, Rev. Mod. Phys. 77, 259 (2005).
  
  
  \bibitem{tasaki} 
H. Tasaki, Long-range order, "tower" of states and symmetry breaking in lattice quantum systems, J. of Stat. Phys. 174, 735-761 (2019).

 \bibitem{liang}
  S. Liang, B. Doucot, and P. W. Anderson, Some new variational resonating-valence-bond-type wave functions for the spin-1/2 antiferromagnetic Heisenberg model on a square lattice, 
  Phys. Rev. Lett. 61, 365, (1988).
  
   \bibitem{feynman} Feynman Lectures on Physics, Vol. 3 Ch 12, R.P. Feynman, R. B. Leighton and M. Sands, 1965, observation attributed to Dirac.  
   
 \bibitem{tomalia} 
 D. Tomalia, Dendrimer molecules, Scientific American  Vol. 273, issue 273 May (1995).
 
 \bibitem{hervet} 
 P. G. DeGennes and H. J. Hervet,  Statistics of Starburst Polymers, J. de Physique Lett. 44, 351 (1983).
 
 \textcolor{red} {
 \bibitem{bentsen} 
 G. Bentsen, T. Hashizume, A. Buyskikh, E. David, A. Daley, S. Gubser, and M. Schleier-Smith, Tree like interaction and fast scramblers with cold atoms  
 Phys. Rev. Lett. 123, 13061 (2019).   For a popular exposition, see
  A. Becker, One lab's quest to build space-time out of quantum particles, Quanta  Magazine, Sep. 7, 2021.  }

 
 \bibitem{song} 
H. F. Song, N. Laflorencie, S. Rachel, and K. LeHur, Entanglement entropy of the two-dimensional Heisenberg anti ferromagnet, Phys. Rev. B 83, 224410, (2011).

 \textcolor{red} {
\bibitem{Met} 
M. A. Metlitski, and T. Grover, Entanglement Entropy of Systems with Spontaneously Broken Continuous Symmetry, arXiv: 1112.5166. }

\bibitem{luitz} D. J. Luitz, X. Plat, F. Alet, and N. Laflorencie, Universal logarithmic corrections to entanglement entropies in two dimensions with spontaneously broken continuous symmetries, Phys. Rev. B 91, 155145 (2015).

\bibitem{takahashi}
M. Takahashi, Modified spin-wave theory of a square-lattice antiferromagnet, Phys. Rev. B 40, 2494 (1989).

\bibitem{hirsch}
J. E. Hirsch and S. Tang, Spin-wave theory of the quantum antiferromagnet with unbroken sublattice symmetry, Phys. Rev. B 40, 4769 (1989).


  \bibitem{lieb2}
  E. H. Lieb, Two theorems on the Hubbard model, Phys. Rev. Lett. 62, 1201 (1989).
  
  \bibitem{mattis} D. C.  Mattis, Ground-State Symmetry in XY model of Magnetism, Phys. Rev. Lett. 42, 1503 (1979).

  
  \bibitem{gomez} G. Gomez-Santos and J. D. Joannopoulos,  Application of spin-wave theory to the ground state of XY quantum Hamiltonians, Phys. Rev. B 36, 8707 (1987).


\bibitem{bohdan} B. Kulchytskyy, C. M. Herdman, S. Inglis, and R.  G. Melko,  Detecting Goldstone Modes with Entanglement Entropy, Phys. Rev. B 92,  115146 (2015).


\bibitem{bauer}  Dag-Vidar Bauer and J. O. Fjærestad, Spin-wave study of entanglement and Renyi entropy for coplanar and collinear magnetic orders in two-dimensional quantum Heisenberg anti ferromagnets,  Phys. Rev. B 101, 195124 (2020).


\bibitem{berthiere} Clément Berthiere and William Witczak-Krempa, Entanglement of Skeletal Regions, Phys. Rev. Lett. 128, 240502 (2022).

\bibitem{chan}  \textcolor{red} {Seughoon Lee et al. ,  "Is There Evidence of Exponential Quantum Advantage in Quantum Chemistry?"  arXiv: 2208.02199} ;  Garnet Chan, Simons Institute, Berkeley,  talk of the same title , Apr. 2022.
 


 \bibitem{schreiber} 
  Y. Schrieber, and R. Berkovits, Entanglement entropy on the Cayley Tree, 
  Journal of Statistical Mechanics: Theory and Experiment (2016) (8) 083104.


  \bibitem{lieb}
  E. H. Lieb, and D. Mattis, Ordering energy levels of interacting spin systems , 
  J. Math. Phys. 3, 749, (1962).
  
  
  \bibitem{alet}
  F. Alet, S. Capponi, N. Laflorencie, and M. Mambrini, Valence bond entanglement entropy,
  Phys. Rev. Lett. 99, 117204, (2007).
  
   \bibitem{chhajlany}
  R. W. Chhajlany, P. Tomczak, and A. Wojcik, Topological estimator of  block entanglement for Heisenberg anti ferromagnets,
  Phys. Rev. Lett. 99, 167204, (2007).
  
   \bibitem{kallin5}
  A. B. Kallin, I. Gonzalez, M. B. Hastings and R. G. Melko, Valence bond and von Neumann entanglement entropies in Heisenberg Ladders,
  Phys. Rev. Lett. 103, 117203, (2009).
  
  \bibitem{peschel}
I. Peschel and V. Eisler, Reduced density matrices and entanglement entropy in free lattice models,  J. Phys. A: Math. Theor. 42, 504003 (2009).

\bibitem{Changlani}
H. J. Changlani, S. Ghosh, C.  L. Henley, and A. M. Lauchl, Heisenberg antiferromagnet on Cayley trees: Low-energy spectrum and even/odd site imbalance, Phys. Rev. B 87, 085107 (2013).


\bibitem{Henley}
H. J. Changlani,  S. Ghosh, S. Pujari, and C. L. Henley, Emergent Spin Excitations in a Bethe Lattice at Percolation,  Phy. Rev. Lett 111, 157201 (2013).










      


  

  

  

  
 
  
  
   
 
    
  
   

  
    
  
  









 
         

 
\end{references}
\end{document}